\documentclass[sts,preprint]{imsart}

\RequirePackage{amsthm,amsmath,amsfonts,amssymb}
\RequirePackage[authoryear,round]{natbib}
\RequirePackage{booktabs,tabularx,array,graphicx}
\RequirePackage[colorlinks,citecolor=blue,urlcolor=blue]{hyperref}

\startlocaldefs
\numberwithin{equation}{section}
\theoremstyle{plain}
\newtheorem{proposition}{Proposition}[section]
\theoremstyle{definition}

\newtheorem{example}{Example}[section]

\newcolumntype{Y}{>{\raggedright\arraybackslash}X}
\newcommand{\E}{\mathbb{E}}
\newcommand{\Pp}{\mathbb{P}}
\newcommand{\R}{\mathbb{R}}

\newcommand{\MMD}{\operatorname{MMD}}
\newcommand{\argmax}{\mathop{\rm arg\,max}}

\endlocaldefs

\begin{document}

\begin{frontmatter}

\title{Nonparametric Change-Point Detection for Modern Data: Representations and Statistical Guarantees}
\runtitle{Nonparametric Change-Point Detection for Modern Data}

\begin{aug}
\author[A]{Hoseung Song}
\address[A]{Hoseung Song is an Assistant Professor, Department of Industrial and Systems Engineering, KAIST, Daejeon, Republic of Korea (e-mail: hoseung@kaist.ac.kr)}
\end{aug}

\begin{abstract}
	Nonparametric change-point methods for high-dimensional and non-Euclidean data, including networks and covariance matrices, compare observations through a representation: coordinates, kernels, distances, similarity graphs, ranks, or learned scores. These representations affect which changes a method can detect and what the reported evidence means. We review retrospective methods by relating their representations to the population discrepancies targeted by their scores, the search over change times, the calibration of that search, and the resulting statistical guarantee. We distinguish population identification, detection power, scan-level error control, and localization, and examine these guarantees for multiple changes, dependent sequences, objects estimated from replicates, and online monitoring. Simulation experiments illustrate these distinctions: two scores on the same nearest-neighbor graph have similar power but give different location estimates. These comparisons guide the choice of a representation that preserves the change of scientific interest and a procedure whose guarantee supports the intended inference.
\end{abstract}

\begin{keyword}
\kwd{high-dimensional data}
\kwd{non-Euclidean data}
\kwd{permutation tests}
\kwd{kernel and energy statistics}
\kwd{graph-based tests}
\kwd{localization}
\end{keyword}

\end{frontmatter}

%%%%%%%%%%%%%%%%%%%%%%%%%%%%%%%%%%%%%%%%%%%%%%%%%%%%%%%%%%%%%%%%%%%%%%%%%%%%%%%%%%%%%%%%%%%%%%%%%%%%%%%%%%%%%%%%%%%%%%%%%%%%%%%%%%%%%%%%%%%%%%%%

\section{Why representation matters}
\label{sec:introduction}

Over an academic term, do different pairs of people come into contact, or does the number of such pairs change? In an analysis of data from the MIT Reality Mining study \citep{eagle2006reality},  \citet{dubeyzheng2026profiles} represented each of 232 days by a Bluetooth proximity network among the same 96 participants. Their scan, based on Frobenius distances between graph Laplacians, placed a change on December 15, 2004, during finals week. This distance responds to which pairs are in contact, so it bears on the first question. Counting the distinct pairs in proximity each day would instead answer the second question but not the first: two days can have identical counts even when their sets of pairs are disjoint. The academic calendar suggests an interpretation of the estimated date, but judging how surprising the largest scan value is requires a model for day-to-day variation, including any weekly pattern.

This example illustrates two general questions that apply equally to vectors, curves, matrices, and networks. What features of the observations must be retained for the change to remain detectable? How should the evidence be calibrated when the change time is selected by searching over many candidates? We call the form in which observations enter a method (e.g., coordinates, a kernel, a distance, a similarity graph, ranks, or a learned score) its representation. Even when a representation preserves the change at the population level, a finite sample may carry little evidence for it. Searching over dates poses a separate problem: a threshold calibrated for a prespecified date may no longer control the type I error rate when applied to the scan maximum.

We organize the methods through the sequence in (\ref{eq:pipeline}): a representation of the observations, the population discrepancy that its score estimates, a search over candidate change times, a calibration of that search, and the guarantee that results.
\begin{equation}
\begin{aligned}
\text{representation}
&\longrightarrow \text{population discrepancy}\\
&\longrightarrow \text{temporal search}\\
&\longrightarrow \text{calibration}
\longrightarrow \text{guarantee}.
\end{aligned}
\label{eq:pipeline}
\end{equation}

%%%%%%%%%%%%%%%%%%%%%%%%%%%%%%%%%%%%%%%%%%%%%

\subsection{Scope, selection, and relation to existing reviews}

Our focus is retrospective detection of distributional changes in high-dimensional and non-Euclidean observations. Classical accounts supply the foundations for nonparametric change-point analysis \citep{brodsky1993,csorgo1997}. Existing reviews organize the field around multiple changes \citep{niu2016}, offline costs and searches \citep{truong2020}, high-dimensional inference \citep{liu2022review}, similarity graphs \citep{chen2023review}, and data segmentation \citep{cho2024review}. A recent overview covers change-point methods across applications \citep{li2026review}. 

In this paper, we compare these literatures where they address the same inferential question, and we trace how the answers and their guarantees change when the same observations are represented differently. To this end, we include a method when it contributes a distinct representation, a way to calibrate a search over candidate change times, or a guarantee for a specified inferential target. The main comparison starts with independent observations and a single possible change. We then consider which conclusions carry over to segmentation, dependent or replicated observations, and online monitoring.

%%%%%%%%%%%%%%%%%%%%%%%%%%%%%%%%%%%%%%%%%%%%%

\subsection{A baseline model and the meanings of ``nonparametric''}  \label{subsec:nonpara}

Let $X_1,\ldots,X_n$ be observations taking values in a measurable space $\mathcal{X}$.  In the at-most-one-change (AMOC) model, we consider
\begin{equation}
X_i\sim
\begin{cases}
P,&i\leq \tau,\\
Q,&i>\tau,
\end{cases}
\qquad 1\leq \tau<n,
\label{eq:amoc}
\end{equation}
and the null hypothesis of no change is $H_0:P=Q$.  Unless stated otherwise, $X_1,\ldots,X_n$ are mutually independent, so under $H_0$ they are independent and identically distributed with common distribution $P$; Section~\ref{sec:extensions} relaxes independence. In the multiple-change model, we consider $0=\eta_0<\eta_1<\cdots<\eta_K<\eta_{K+1}=n$ and $X_i\sim P_j$ for $\eta_{j-1}<i\le\eta_j$, $j=1,\ldots,K+1$, with $P_j\neq P_{j+1}$ for $j=1,\ldots,K$. Detectability depends jointly on the minimum segment length $\min_{1\le j\le K+1}(\eta_j-\eta_{j-1})$ and on a signal size measured through the chosen representation.

Strictly speaking, \emph{nonparametric} describes a model that does not restrict the segment distributions to a finite-dimensional parametric family. In classical usage the word also refers to \emph{distribution-free} procedures, whose test statistic has the same null distribution under every data-generating distribution in the null class. Choosing the critical value from this common null distribution, so that the type I error probability is at most $\alpha$, yields a \emph{finite-sample valid} level-$\alpha$ test throughout the null class. An \emph{asymptotically distribution-free} statistic has only a common limiting null distribution, so the corresponding test need not be finite-sample valid.

Two additional properties should be distinguished from nonparametric modeling and distribution-freeness: the ability to identify arbitrary distributional changes and validity under dependence. A discrepancy is \emph{omnibus} on a class of distributions if, for any $P$ and $Q$ in that class, it vanishes only when $P=Q$. For continuous univariate data, the rank statistic of \cite{pettitt1979} is distribution-free but responds mainly to changes in location, not to arbitrary changes in distribution. Conversely, a statistic built on an omnibus discrepancy may have a null distribution that depends on the unknown distribution and therefore require additional calibration. Thus, distribution-freeness and the omnibus property need not hold together. The second property concerns dependence: a procedure is \emph{robust to dependence} over a stated class of processes if its validity holds for every process in that class. 

The remainder of this paper is organized as follows. Sections~\ref{sec:framework} and~\ref{sec:guarantees} develop the representation and guarantee sides of the sequence~\eqref{eq:pipeline}. Section~\ref{sec:diagnostics} turns them into questions to settle when choosing a method, and Section~\ref{sec:protocol} compares implementations on common simulated data. Section~\ref{sec:extensions} considers other observation units (dependent, repeated, and object-valued observations) and online monitoring; Section~\ref{sec:open} discusses open questions, and we conclude with a discussion in Section \ref{sec:discussion}. Code to reproduce all simulations and figures is available at \url{https://github.com/hoseungs/nonparametric-cpd}.

%%%%%%%%%%%%%%%%%%%%%%%%%%%%%%%%%%%%%%%%%%%%%%%%%%%%%%%%%%%%%%%%%%%%%%%%%%%%%%%%%%%%%%%%%%%%%%%%%%%%%%%%%%%%%%%%%%%%%%%%%%%%%%%%%%%%%%%%%%%%%%%%

\section{A representation-to-guarantee framework}
\label{sec:framework}

We first introduce notation for the representation, the population discrepancy, and the temporal search in \eqref{eq:pipeline}; Section \ref{sec:guarantees} treats calibration and guarantees. For a segment $(s,e]=\{s+1,\ldots,e\}$, let $\mathcal R_{s,e}(X_{s+1:e})$ denote a representation of the observations in that segment. At a candidate split $t\in\{s+1,\ldots,e-1\}$, the local score is
\begin{equation}
	D_{s,e}(t)=\Psi_t\!\left\{\mathcal R_{s,e}(X_{s+1:e})\right\},
	\label{eq:localstat}
\end{equation}
where $\Psi_t$ uses the representation to compare the blocks $(s,t]$ and $(t,e]$. The local score may be standardized using its null mean and standard deviation. The representation may consist of pointwise features or quantities computed jointly from the pooled observations, such as similarity graphs, distance matrices, and pooled ranks. If a learned score is fitted using the block labels induced by $t$ (the split labels), the fitting step can be included in $\Psi_t$.

Let $\mathcal T_{s,e}$ be the set of admissible splits, typically restricted so that both blocks meet a minimum size. An AMOC scan takes the maximum local score as its test statistic and a maximizing split as the change-point estimate,
\begin{equation}
	M_{s,e}=\max_{t\in\mathcal T_{s,e}}D_{s,e}(t),
	\qquad
	\widehat\tau\in\argmax_{t\in\mathcal T_{s,e}}D_{s,e}(t).
	\label{eq:scan}
\end{equation}
When the segment is the whole sample, $(s, e] = (0, n]$, we write $S_n(t) = D_{0,n}(t)$, $\mathcal T_n = \mathcal T_{0,n}$, and $M_n = M_{0,n}$. For multiple changes, segmentation methods may scan a collection of intervals, apply the scan recursively, or optimize a penalized sum of segment costs. In univariate sequences, for example, these costs can be constructed from the empirical distribution function \citep{zou2014,haynes2017}.

%%%%%%%%%%%%%%%%%%%%%%%%%%%%%%%%%%%%%%%%%%%%%

\subsection{Representation blindness and data-driven representations}

The next statement formalizes why representation choice cannot be separated from power.

\begin{proposition}[Representation blindness]
	\label{prop:blindness}
	Let $R:\mathcal X\to\mathcal Z$ be a fixed measurable map, and write $R_{\#}P$ for the distribution of $R(X)$ when $X\sim P$. If $R_{\#}P=R_{\#}Q$, then any test or change-point estimator based only on $R(X_1),\ldots,R(X_n)$ has the same distribution under the independent AMOC model~\eqref{eq:amoc} as under the no-change model $X_1,\ldots,X_n\stackrel{i.i.d.}{\sim}P$. In particular, any such level-$\alpha$ test has power at most $\alpha$ against this change.
\end{proposition}

\begin{proof}
	The pointwise map $R$ preserves independence. Under the change model, the represented observations have distribution $R_{\#}P$ up to the change and $R_{\#}Q$ after it. Since these distributions coincide, the represented sequence has joint distribution $(R_{\#}P)^{\otimes n}$ under both models.
\end{proof}

Proposition \ref{prop:blindness} concerns pointwise representations, which apply the same map $R$ to each observation, so that $\mathcal R_{s,e}(X_{s+1:e}) = (R(X_{s+1}), \ldots, R(X_e))$ for every segment $(s, e]$. For a pooled graph or distance matrix, the same conclusion holds if the entire time-indexed representation has the same distribution under the change and no-change models. Equality of the entrywise marginal distributions does not suffice: for a distance matrix, these can agree even when the joint distributions differ, and the change can then remain detectable.

Proposition \ref{prop:blindness} assumes a fixed map. When the representation is instead fitted using the split labels, calibration must account for the fitting step. A resampling procedure can do so by repeating the fit within each resampled scan. When the training data are independent of the observations being scanned, calibration can be justified conditional on the fitted score. 

Searching over representations adds another source of multiplicity to the temporal scan. Let $\mathcal L_n$ be the library of candidate representations or tuning choices, such as coordinates, projections, sparsity levels, kernel bandwidths (Section \ref{subsec:kernel}), or graph densities (Section \ref{subsec:graph}). If $S_{r,n}(t)$ is the local score for choice $r \in \mathcal L_n$ at an admissible split $t \in \mathcal T_n$, the joint search may take the form
\begin{equation} 
	\max_{r \in \mathcal L_n} \max_{t \in \mathcal T_n} S_{r,n}(t).
	\label{eq:double-scan}
\end{equation}
Calibration for this maximum must account for both searches. For a prespecified library, Bonferroni correction can be applied to valid full-scan $p$-values for the individual candidates; it remains valid but can be conservative, because it ignores dependence among the candidate scans. Repeating the full search in each valid null resample, or using a simultaneous null approximation, accounts for that dependence. If time labels are used to prune the library, the resampling must repeat the pruning within each resample. 

Sections \ref{subsec:coordinate}--\ref{subsec:object} examine the main representations in turn. Table \ref{tab:representations} summarizes the signal each retains, where it can lose power, and what its calibration requires.
\begin{table*}[t]
\caption{Signal retained by each representation, where it can lose power, and what its calibration requires.}
\label{tab:representations}
\centering
\footnotesize
\begin{tabularx}{\textwidth}{@{}>{\raggedright\arraybackslash}p{.16\textwidth}YYY@{}}
\toprule
Representation & Signal retained & Where power is lost & Calibration requirement \\
\midrule
Coordinate contrasts & A maximum captures a few large marginal shifts. A sum can accumulate many small shifts. & The maximum pays for searching many coordinates. Unchanged coordinates add noise to a sum. & Account for dependence across coordinates and for the search over time. \\
\specialrule{0.3pt}{1.1pt}{1.1pt}
Projection or subspace & The part of the change along the retained directions & A mean shift orthogonal to the subspace vanishes. Short segments can give unstable directions. & Include data-driven direction selection in the null calculation, or train independently. \\
\specialrule{0.3pt}{1.1pt}{1.1pt}
Kernel & Difference between kernel mean embeddings & Distances can concentrate as dimension grows, leaving little separation at the chosen bandwidth. & Calibrate over the splits and bandwidths searched. \\
\specialrule{0.3pt}{1.1pt}{1.1pt}
Distance or energy & Between-block versus within-block distances. Under finite first moments, strong negative type identifies any change in distribution. & An unsuitable metric can miss a change. Unchanged coordinates can dominate distances as dimension grows. & For permutation calibration, use units that are exchangeable under the null and calibrate the full scan. \\
\specialrule{0.3pt}{1.1pt}{1.1pt}
Similarity graph & Changes in edge counts within or between candidate blocks & Hubs or cancellation between edge counts can obscure a change near its true location. & Permutation requires exchangeable labels. Gaussian scan approximations require graph regularity. \\
\specialrule{0.3pt}{1.1pt}{1.1pt}
Ranks & Changes in the ordering induced by a score & The ordering can discard direction or magnitude. & Justify the null distribution of the ranks, and account for ties and the temporal search. \\
\specialrule{0.3pt}{1.1pt}{1.1pt}
Classifier or learned score & Separation of the two temporal blocks by their features & Fitting on short blocks can obscure weak interactions or exaggerate separation. & Account for time labels used in training and for the choice of split. \\
\specialrule{0.3pt}{1.1pt}{1.1pt}
Object-specific features & Changes in a chosen geometric feature, such as a Fr\'echet mean or a spectrum & A feature invariant to the change misses it, as when covariance matrices with different orientations share a spectrum. & Object-level permutation requires whole objects to be exchangeable under the null. \\
\bottomrule
\end{tabularx}
\end{table*}

%%%%%%%%%%%%%%%%%%%%%%%%%%%%%%%%%%%%%%%%%%%%%

\subsection{Coordinatewise contrasts, aggregation, and projections} \label{subsec:coordinate}

For observations $X_i\in\R^p$, one approach is to compare the two blocks coordinate by coordinate. Let $P^{(j)}$ and $Q^{(j)}$ be the marginal distributions of the $j$-th coordinate. For a univariate discrepancy $d$, define
\begin{equation}
	\Delta_j(P,Q)=d(P^{(j)},Q^{(j)}),\quad j=1,\ldots,p.
	\label{eq:coordinate-signal}
\end{equation}
A maximum of the estimated discrepancies can capture a change concentrated in a few coordinates, with a multiplicity cost for searching across coordinates. A sum or quadratic form can accumulate many small discrepancies, along with noise from unchanged coordinates, and may require covariance estimation. Multiscale local tests can adapt to sparse high-dimensional mean changes \citep{pilliat2023}, and distance-based statistics can target non-sparse alternatives \citep{drikvandi2025}. The relative performance of these approaches depends on the sparsity and magnitude of the change, dependence across coordinates, and segment length. If $P^{(j)}=Q^{(j)}$ for every $j$, all discrepancies in \eqref{eq:coordinate-signal} are zero, but the joint distribution of their estimators can still change with the dependence among coordinates. A coordinatewise aggregate can therefore respond to a change in dependence without being omnibus for the joint distribution (Example \ref{ex4.1}). A discrepancy computed from the full observation vectors can target changes in the joint distribution directly.

Projection is a second way to reduce dimension. For a mean shift $\delta \in\R^p$, projection onto a unit vector $u$ gives a projected mean shift of $u^\top\delta$. Its magnitude is $\|\delta\|$ when $u$ is parallel to $\delta$ and zero when $u$ is orthogonal to $\delta$. For a general distributional change, Proposition~\ref{prop:blindness} applies when, under a fixed projection, the projected observations have the same distribution before and after the change. For an unknown sparse mean shift, \citet{wang2018inspect} estimate a projection direction from the $p\times(n-1)$ matrix of coordinatewise cumulative sum (CUSUM) statistics and scan the projected sequence to estimate the change location. Their localization guarantees depend on how accurately the direction is estimated. \citet{yu2026kme} use corrected kernel principal components to estimate a subspace that, under their assumptions, preserves the population number and locations of changes.

%%%%%%%%%%%%%%%%%%%%%%%%%%%%%%%%%%%%%%%%%%%%%

\subsection{Kernels: mean embeddings and their geometry} \label{subsec:kernel}

Let $k$ be a positive-definite kernel with reproducing kernel Hilbert space $\mathcal H_k$ and feature map $\phi$. Let $X,X'\sim P$ and $Y,Y'\sim Q$, with all four variables mutually independent. Assume that the kernel mean embeddings exist; for a strongly measurable feature map, a sufficient condition is $\E\sqrt{k(X,X)}+\E\sqrt{k(Y,Y)}<\infty$. The squared maximum mean discrepancy (MMD) is
\begin{equation}
	\begin{aligned}
		\MMD_k^2(P,Q)
		&=\left\|\E_P\phi(X)-\E_Q\phi(Y)\right\|_{\mathcal H_k}^2\\
		&=\E k(X,X')+\E k(Y,Y')\\
		&\quad-2\E k(X,Y).
	\end{aligned}
	\label{eq:mmd}
\end{equation}

If $k$ is characteristic on a class of distributions, then $\MMD_k(P,Q)=0$ if and only if $P=Q$ within that class \citep{gretton2012}. At a candidate split, biased or unbiased estimators of $\MMD_k^2(P,Q)$ can serve as local two-sample scores. Windowed procedures and scans may standardize kernel scores at candidate times \citep{li2015mstat,song2024cp}. Kernel segment costs also support dynamic programming and model selection \citep{harchaoui2007,harchaoui2009,garreau2018,celisse2018,arlot2019}.

A characteristic kernel guarantees identification, not power. Finite-sample power also depends on the size of the population discrepancy relative to sampling variability. In high dimensions, pairwise distances can concentrate, and the bandwidth of a radial kernel can affect sensitivity to changes in moments of different orders. High-dimensional analyses of kernel two-sample tests describe how these effects depend on dimension and sample size \citep{yan2023}. Generalized kernel statistics can combine components sensitive to different changes \citep{song2024gkt}. Computational approximations such as block averaging can reduce cost but can alter sensitivity to particular changes \citep{song2026fast}. In a change-point scan, the variability of the score also depends on the two block sizes, so a normalization suited to balanced splits may be less stable near the ends of a segment, where one block is short.

%%%%%%%%%%%%%%%%%%%%%%%%%%%%%%%%%%%%%%%%%%%%%

\subsection{Distances and energy statistics}

Let $\rho$ be a metric or semimetric on the observation space. With $X,X',Y,Y'$ as in Section \ref{subsec:kernel} and assuming the distance expectations below are finite, the population energy distance is defined as
\begin{equation}
	\mathcal E(P,Q)=2\E\rho(X,Y)-\E\rho(X,X')-\E\rho(Y,Y').
	\label{eq:energy}
\end{equation}

A metric or semimetric $\rho$ has negative type if
\[
\sum_{i,j=1}^m a_i a_j \rho(x_i,x_j) \leq 0
\]
for every finite collection of points $x_1,\ldots,x_m$ and real coefficients $a_1,\ldots,a_m$ satisfying $\sum_{i=1}^m a_i=0$ \citep{sejdinovic2013}. This condition ensures that $\mathcal E(P,Q)\geq 0$. We say that $\rho$ has strong negative type if, in addition, $\mathcal E(P,Q)=0$ only when $P=Q$ within the class of distributions with finite first $\rho$-moments, that is, $\E\rho(X,o)<\infty$ for some $o$ \citep{lyons2013distance}. \citet{matteson2014} use empirical energy contrasts in a divisive procedure for multiple changes.

For a semimetric $\rho$ of negative type and a fixed anchor $o$, the associated positive-definite kernel is
\[
k_o(x,y)=\{\rho(x,o)+\rho(y,o)-\rho(x,y)\}/2.
\]
Substituting $k_o$ into \eqref{eq:mmd} cancels the anchor terms under the first-moment condition and gives $\mathcal E(P,Q)=2\MMD_{k_o}^2(P,Q)$ \citep{sejdinovic2013,szekely2013}. The same identity holds for the empirical contrasts at each split when the within-block averages in both contrasts use the same convention for including or excluding diagonal terms. Taking the maximum over identical candidate splits and weights preserves the factor of two. The two scans therefore have identical rejection regions when the energy critical value is twice the squared-MMD critical value, as occurs when both are calibrated using the same null resamples. For example, the MMD scan in Section~\ref{sec:protocol} has an equivalent energy scan based on the semimetric $\rho_k(x,y)=2\{1-k(x,y)\}$, where $k$ is the Gaussian kernel at the chosen bandwidth.

%%%%%%%%%%%%%%%%%%%%%%%%%%%%%%%%%%%%%%%%%%%%%

\subsection{Similarity graphs: sparse relational summaries} \label{subsec:graph}

Graph-based methods build a similarity graph $G$ on the pooled observations, with one vertex per observation and edges joining pairs selected according to a chosen dissimilarity. Common choices include a minimum spanning tree, a matching, and a $k$-nearest-neighbor graph. At each candidate split, the vertices are labeled according to their block. The original edge-count statistic looks for fewer edges between the two blocks than expected under the permutation null \citep{friedman1979,rosenbaum2005}. Generalized edge-count statistics use the two within-block edge counts and can also respond to an imbalance between them, as can occur under a change in scale \citep{chenfriedman2017}.

A sparse graph summarizes the pairwise dissimilarities with a smaller set of edges for the scan to process. In a $k$-nearest-neighbor graph, $k$ sets how local the comparison is. Small neighborhoods can omit informative links, while larger neighborhoods can dilute contrasts that are concentrated locally. Ties can affect which edges are selected, and hubs or disconnected components can affect how the edges contribute to the score.

When graph construction treats the pooled observations symmetrically and does not use their time order, permutation calibration can hold the graph fixed and permute the time labels, provided that the observations are exchangeable under the null. The same symmetry requirement applies to any rule used to break distance ties. The permutation means and variances of the edge counts have closed forms, which make analytic approximations to the null distribution of the scan maximum possible; these Gaussian-process approximations require regularity conditions on the graph sequence \citep{chen2015,chu2019}. Extensions handle the ambiguity in graph construction that arises when some observations take identical values \citep{song2022repeated}.

%%%%%%%%%%%%%%%%%%%%%%%%%%%%%%%%%%%%%%%%%%%%%

\subsection{Ranks and learned scores} \label{subsec:rank}

Under the no-change null, the ranks of independent observations from a common continuous univariate distribution are uniformly distributed over all permutations \citep{pettitt1979}. For multivariate data, ranks can be constructed from projections, center-outward maps \citep{hallin2021distribution}, or similarity graphs. Rank-based homogeneity tests extend this approach to multivariate comparisons \citep{lungyutfong2015}, and graph-induced ranks can yield scans with asymptotically distribution-free calibration \citep{zhou2025rank}. Ranking can simplify null calibration, at the cost of discarding some magnitude information.

Recent work shows that data-adaptive scoring need not forfeit finite-sample calibration. For example, the ART procedure of \cite{cui2026art} fits a score function symmetrically to the pooled sample, so that the fitted function does not depend on the order of the observations, and assigns each observation its scalar score. Permuting the observations therefore permutes their scores in the same way. Under independent sampling from a common distribution, the fitted scores are exchangeable even when the common fit makes them dependent. With randomized tie handling, their ranks are uniformly distributed over permutations. A prespecified rank scan can thus be calibrated in finite samples by permuting the ranks of the fitted scores. The multiscale extension of ART uses local ranks on data-independent intervals to support multiple-change testing, localization regions, and post-detection error control. Validity, however, does not imply power: detection power and localization accuracy depend on the information retained by the scores, the strength of the changes, and their spacing.

Classifier methods turn the candidate split into a binary label and ask how well observations from the left and right blocks can be separated. Examples include the random-forest classifier in \texttt{changeforest} \citep{londschien2023} and discrepancies based on the area under the ROC curve \citep{kanrar2025}. With equal class priors and zero-one loss, the Bayes error is $\{1-\operatorname{TV}(P,Q)\}/2$, where $\operatorname{TV}(P,Q)=\sup_A |P(A)-Q(A)|$ over measurable sets $A$. An optimal classifier therefore targets total variation: its accuracy exceeds random guessing by $\operatorname{TV}(P,Q)/2$.

When class priors reflect unequal block sizes, always predicting the larger block can yield high accuracy near the ends of a scan, so raw accuracy should be compared with the majority-class rate rather than with 1/2. Flexible classifiers fitted on short blocks can overfit; sample splitting or cross-fitting assesses separation on held-out observations. After a search over splits, valid inference requires calibrating the full scan, for example by a permutation test that repeats the fit within each permuted scan.

CUSUM-Net learns a scalar-valued function, called a critic, by maximizing a CUSUM objective that aggregates contrasts over candidate splits \citep{li2026cusumnet}. A classical CUSUM scan then estimates the change location from the resulting sequence of scores. The candidate score functions define an integral probability metric, which measures the largest difference in expectations over that function class. Under smoothness and signal assumptions, the authors derive localization bounds from the difference between the CUSUM objective values at the fitted and optimal critics.

%%%%%%%%%%%%%%%%%%%%%%%%%%%%%%%%%%%%%%%%%%%%%

\subsection{Object-specific structure} \label{subsec:object}

For object-valued observations with finite expected squared distance, a Fr\'echet mean, when it exists, minimizes the expected squared distance to the random object. The Fr\'echet variance is the value of this minimum \citep{dubey2020}. Distinct distributions can share both summaries, so discrepancies that compare only these summaries do not identify every distributional change.

For time series in a separable metric space of strong negative type, such as positive-definite matrices with the log-Euclidean metric, \citet{zhang2026object} use a distance-induced Hilbert-space embedding. The mean of this embedding identifies the marginal distribution within the class of distributions with finite first distance moments. They estimate a projection direction from an initial and a final block of the sequence, which are excluded from the scan, and apply a self-normalized scan to the projected scores of the middle block. The asymptotic calibration of the scan requires additional moment and weak-dependence conditions. 

For sequences of independent Bernoulli networks on a common vertex set, \citet{wang2021network} derive localization rates for changes in the matrix of edge probabilities under their sparsity, rank, and signal conditions. Section~\ref{sec:extensions} discusses the sampling units underlying these analyses.

\begin{table*}[t]
\caption{Selected recent methods organized by the steps of (\ref{eq:pipeline}): representation and target, temporal search and principal guarantee, and conditions and limitations.}
\label{tab:recent-methods}
\centering
\scriptsize
\begin{tabularx}{\textwidth}{@{}p{0.15\textwidth}p{0.23\textwidth}p{0.25\textwidth}Y@{}}
\toprule
Method & Representation and target & Temporal search and principal guarantee & Conditions and limitations \\
\midrule
ART \citep{cui2026art} & Symmetric pooled-sample score followed by local ranks & AMOC and multiscale aggregation; finite-sample distribution-free error control under independence and symmetry & Validity does not make an arbitrary score powerful; order-adaptive fitting can break symmetry \\
\specialrule{0.3pt}{1.1pt}{1.1pt}
CUSUM-Net \citep{li2026cusumnet} & Neural critic estimating an integral probability metric & Aggregate CUSUM learning and localization; excess-risk-to-localization theory & Theory uses smoothness, architecture, intrinsic dimension, signal, and spacing assumptions \\
\specialrule{0.3pt}{1.1pt}{1.1pt}
Kernel mean embedding deviation subspace \citep{yu2026kme} & Corrected kernel principal subspace preserving embedding deviations & Dimension reduction before segmentation; preservation of change number and locations under their conditions & Preservation is conditional on the subspace model and corrected construction \\
\specialrule{0.3pt}{1.1pt}{1.1pt}
NP-MOJO \citep{mcgonigle2025} & Joint characteristic functions of observations and lags & Moving-window multiscale segmentation; consistency for marginal and serial-dependence changes & Lag set, window scale, fixed dimension, and short-range dependence define the target and regime \\
\specialrule{0.3pt}{1.1pt}{1.1pt}
Object-valued scan \citep{zhang2026object} & Distance-induced Hilbert embedding for changes in the marginal distribution & Self-normalized AMOC test with a pivotal null limit; wild binary segmentation for multiple changes & Strong negative type for identification; moment and weak-dependence conditions for asymptotic inference  \\
\specialrule{0.3pt}{1.1pt}{1.1pt}
Multiscale local-test aggregation \citep{pilliat2023} & A modular family of local homogeneity tests & Bottom-up multiple-change aggregation; no-spurious-change and localization properties on a simultaneous good event & Optimality rates belong to specific local tests and alternative classes, not to aggregation alone \\
\specialrule{0.3pt}{1.1pt}{1.1pt}
Conformal localization \citep{dandapanthula2026} & Candidate-wise conformal scores and a matrix of $p$-values & Finite-sample confidence set including the no-change candidate & Width, power, and post-selection coverage require separate analysis \\
\specialrule{0.3pt}{1.1pt}{1.1pt}
E-detectors \citep{shin2024edetectors} & Restarted and mixed e-process evidence & Sequential stopping with nonasymptotic average-run-length control and delay results in specified models & The main guarantee is nonasymptotic control of the average run length (ARL), not of the probability of ever raising a false alarm \\
\bottomrule
\end{tabularx}
\end{table*}

Table~\ref{tab:recent-methods} summarizes the representations, temporal searches, and statistical guarantees of selected methods discussed in Sections \ref{sec:framework}, \ref{sec:guarantees}, and \ref{sec:extensions}. Population identification, detection power, scan-level error control, and localization each require separate arguments, which Section \ref{sec:guarantees} develops.

%%%%%%%%%%%%%%%%%%%%%%%%%%%%%%%%%%%%%%%%%%%%%%%%%%%%%%%%%%%%%%%%%%%%%%%%%%%%%%%%%%%%%%%%%%%%%%%%%%%%%%%%%%%%%%%%%%%%%%%%%%%%%%%%%%%%%%%%%%%%%%%%

\section{Statistical guarantees and their scope}
\label{sec:guarantees}

The scope of a statistical guarantee depends on its target, the class of data-generating processes, and the full procedure, including its temporal search and calibration. Table~\ref{tab:guarantees} summarizes the conditions needed to justify the guarantees considered in this section.

%%%%%%%%%%%%%%%%%%%%%%%%%%%%%%%%%%%%%%%%%%%%%

\subsection{Population identification, detection, and localization}

Let $\Delta_{\mathcal R}(P,Q)$ denote the population discrepancy targeted by a scan built on representation $\mathcal{R}$, such as $\MMD_k(P,Q)$ or $\mathcal E(P,Q)$. Recall from Section \ref{subsec:nonpara} that it is omnibus if $\Delta_{\mathcal R}(P,Q)>0$ for every fixed $P\ne Q$ in its domain. A scan based on local scores $S_n(t)$ rejects when their maximum over the candidate splits $\mathcal T_n$ exceeds the calibrated critical value $c_n$. Write the local score as
\begin{equation}
	S_n(t)=\Gamma_n(t)+\xi_n(t),
	\label{eq:margin}
\end{equation}
where $\Gamma_n(t)$ is a population profile, possibly defined conditional on the representation, and $\xi_n(t)$ is the stochastic error relative to that profile.

Suppose the true change-point $\tau$ belongs to $\mathcal T_n$. On the event $\sup_{t\in\mathcal T_n}|\xi_n(t)|\leq a_n$, the inequality $\Gamma_n(\tau)>c_n+a_n$ suffices for rejection. On the same event, a sufficient condition for every maximizer $\widehat\tau$ to satisfy $|\widehat\tau-\tau|<r_n$ is
\begin{equation}
	\inf_{\substack{t\in\mathcal T_n\\|t-\tau|\geq r_n}}
	\{\Gamma_n(\tau)-\Gamma_n(t)\}>2a_n.
	\label{eq:argmaxmargin}
\end{equation}
For each such $t\in \mathcal{T}_n$ with $|t-\tau|\ge r_n$, the error difference $\xi_n(\tau)-\xi_n(t)$ can reduce the population difference by at most $2a_n$, so $S_n(\tau)>S_n(t)$ and no such $t$ maximizes $S_n$.
 
For example, for a fixed kernel, let $\bar\mu_{a:b}$ be the average of $\E\phi(X_i)$ over $a\leq i\leq b$, and set $\Delta^2=\MMD_k^2(P,Q)$. Replacing the empirical block means $\hat{\mu}_{1:t}$ and $\hat{\mu}_{(t+1):n}$ in the weighted squared embedding contrast $t(n-t)/n \cdot \|\hat{\mu}_{1:t}-\hat{\mu}_{(t+1):n}\|^2$ by their population counterparts gives
\[
\Gamma_n(t)=\frac{t(n-t)}{n}
\left\|\bar\mu_{1:t}-\bar\mu_{(t+1):n}\right\|^2_{\mathcal H_k}.
\]
At $t=\tau$, the profile height is $\Gamma_n(\tau)=\tau(n-\tau)\Delta^2/n$. When $t<\tau$, the right block contains observations from both distributions. When $t>\tau$, the left block does. Substituting the corresponding block averages gives
\[
\Gamma_n(\tau)-\Gamma_n(t)=\Delta^2
\begin{cases}
	(n-\tau)(\tau-t)/(n-t),&t\leq\tau,\\
	\tau(t-\tau)/t,&t>\tau.
\end{cases}
\]
In both cases, $\Gamma_n(\tau) - \Gamma_{n}(t) \ge \lambda_n \Delta^2 |t-\tau |$ with $\lambda_n = \min(\tau, n-\tau)/n$, so (\ref{eq:argmaxmargin}) holds whenever $r_n > 2a_n/(\lambda_n  \Delta^2)$. Therefore, for a given stochastic error bound $a_n$, a smaller $\Delta^2$ gives a weaker localization bound, even when the kernel is characteristic.

%%%%%%%%%%%%%%%%%%%%%%%%%%%%%%%%%%%%%%%%%%%%%

\subsection{Fixed-split and scan-level validity} \label{subsec:fixed}

Suppose that a valid $p$-value $p_t$ is available at each prespecified split $t$. Searching over splits and rejecting when $\min_t p_t\leq \alpha$ can inflate the type I error probability. For $m$ independent uniform $p$-values, this probability is $1-(1-\alpha)^m$. In a change-point scan, the splitwise statistics share observations, and their dependence affects the null distribution of the scan. 

Analytic calibration of $M_{s,e}$ can use a functional limit for the scan process or a direct approximation to the null distribution of its maximum. For example, graph scans use permutation covariance calculations and Gaussian-process limits under graph regularity conditions \citep{chen2015,chu2019}. Kernel scans can be calibrated by analytic approximation or by resampling the full scan \citep{song2024cp}. Near the segment boundaries, one block is short and variance estimates can be unstable, which motivates trimming to $(t-s)/(e-s)\in[\epsilon,1-\epsilon]$.

Let $\mathcal G$ be a finite group of transformations that leave the joint distribution of $X_{1:n}$ under the null invariant. The orbit of $X_{1:n}$ is the set $\{gX_{1:n}:g\in\mathcal G\}$ of all sequences obtained by applying the transformations in $\mathcal G$. Under the null, conditional on the orbit, the observed sequence is equally likely to be $gX_{1:n}$ for each $g\in\mathcal{G}$. The proportion of transformations for which the full scan statistic is at least as large as its observed value gives a valid finite-sample $p$-value. This $p$-value can be conservative because of discreteness and ties. When $B$ transformations drawn at random from $\mathcal{G}$ are used, the Monte Carlo permutation $p$-value $(1+\#\{b:M_b\ge M_{obs}\})/(B+1)$ remains finite-sample valid, where $M_{obs}$ is the observed scan statistic and $M_b$ is its value for the $b$-th transformed sequence.

Each transformed sequence is evaluated using the same full procedure, including maximization over candidate splits and any selection or fitting that uses time labels. An order-invariant structure built from the pooled observations, such as a similarity graph, need not be rebuilt: the graph remains attached to the observations while their time labels are permuted. An equivariant result, such as the vector of scores from a symmetric fit (Section \ref{subsec:rank}), can be permuted along with the observations instead of being recomputed. For a randomized learner, one option is to condition on a seed chosen independently of the data and use it for the observed sequence and all transformations.

Under an i.i.d. no-change null, the observations are exchangeable. With serial dependence, shuffling individual observations can change the joint distribution of the sequence even under the null. Block permutation, block bootstrap, multiplier methods, and long-run variance estimation rely on different assumptions about dependence. Block methods are typically justified asymptotically under conditions on dependence and block length. The finite-sample group argument applies to circular shifts when, under the null, the joint distribution of the observed sequence is invariant under cyclic rotation.

%%%%%%%%%%%%%%%%%%%%%%%%%%%%%%%%%%%%%%%%%%%%%

\subsection{Sample size, null class, and conditioning}

A calibration guarantee specifies the sample-size regime, the null class, and any conditioning. Finite-sample validity bounds the type I error probability by the nominal level at the given sample size. Asymptotic validity imposes the bound on the limit superior of the type I error probability as $n\rightarrow\infty$. A pointwise asymptotic guarantee applies to each fixed distribution in the null class separately. A uniform asymptotic guarantee applies the bound to the supremum of the type I error probability over the null class. A guarantee may be unconditional or conditional on, for example, a fitted representation, a graph, or a permutation orbit.

The simulations in Section \ref{sec:protocol} report type I error rates that are conditional in this sense. A separate set of independent null sequences estimates the type I error rate conditional on the realized critical value. The unconditional type I error rate averages over repeated draws of the calibration set. Evaluating on the calibration set itself would largely reproduce the empirical tail fraction used to choose the critical value.

%%%%%%%%%%%%%%%%%%%%%%%%%%%%%%%%%%%%%%%%%%%%%

\subsection{High-dimensional power and detection boundaries}

Let the dimension $p$ depend on $n$, let $\lambda_n=\min(\tau,n-\tau)/n$ be the fraction of observations in the smaller block, and let $\Delta_n=\Delta_{\mathcal R}(P_n,Q_n)$ measure the signal in the chosen representation. A power statement specifies how these quantities vary with sample size, together with the relevant assumptions on sparsity, tails, dependence, and tuning. A minimax detection boundary is defined for a specified model, class of alternatives, signal measure, and testing risk. An upper bound shows that some test attains the target risk whenever the signal exceeds a given scale. A matching lower bound shows that below that scale, the worst-case risk of every test remains above the target.

Under exchangeability, a permutation test based on the full scan statistic can control the type I error probability in finite samples at any dimension. Power depends on how strongly the chosen representation separates the null and alternative. When $p$ grows with $n$, an asymptotic calibration also requires a null approximation that is valid in that joint regime.

At a fixed split, the level-$\alpha$ critical value for the largest absolute score among $p$ independent standard Gaussian coordinate scores is of order $\sqrt{\log p}$, and the full scan must also pay for the search over splits. The other high-dimensional effects (noise accumulation in sums, distance concentration for radial kernels, hubs in graphs, and unstable classifiers on short blocks) are discussed in Section \ref{sec:framework} and summarized in Table \ref{tab:representations}.

Different methods address different high-dimensional regimes. The distance-based statistics of \cite{drikvandi2025} target non-sparse changes and cover settings with fixed $n$ and growing $p$. Kernel mean embedding deviation subspaces preserve the number and locations of population changes under their conditions \citep{yu2026kme}. CUSUM-Net relates critic estimation and intrinsic dimension to localization \citep{li2026cusumnet}. The signal scale also varies across models. For univariate distributions, \citet{padilla2021} measure the jump by Kolmogorov--Smirnov distance and allow multiple observations at each time point. For multivariate Lipschitz densities, \citet{padilla2022} use the supremum norm of the density difference. Comparing rates therefore requires a common model, signal measure, and inferential target.

%%%%%%%%%%%%%%%%%%%%%%%%%%%%%%%%%%%%%%%%%%%%%

\subsection{Location estimates, confidence sets, and segmentation} \label{subsec:location}

When a single change is present, consistency on the normalized time scale means that $|\widehat\tau-\tau|=o_{\Pp}(n)$. A localization rate $O_{\Pp}(r_n)$ with $r_n=o(n)$ gives a sharper description of the error on the original time scale. Exact recovery requires $\Pp(\widehat\tau=\tau)\to1$. By contrast, an $O_{\Pp}(1)$ location error need not vanish: it can remain nonzero with probability bounded away from zero. Location error among detections can look small when difficult changes are missed. Section~\ref{sec:protocol} reports the miss rate and a loss that penalizes missed changes, alongside location error among detections.

A confidence set quantifies uncertainty about the change location. The procedure of \cite{dandapanthula2026} assigns a $p$-value $q_t$ to each candidate $t\in\{1,\ldots,n\}$, with $t=n$ representing no change. Unlike the fixed-split $p$-values of Section \ref{subsec:fixed}, $q_t$ tests whether the data are compatible with a change at $t$: under this candidate null, the two blocks are each exchangeable and independent of one another. Inverting these tests gives
\begin{equation}
	\mathcal C_{1-\alpha}
	=\{t\in\{1,\ldots,n\}:q_t>\alpha\}.
	\label{eq:conformal-set}
\end{equation}
If $q_t$ is valid at the true location, then $\Pp (\tau\in \mathcal{C}_{1-\alpha}) \ge 1-\alpha$; no multiplicity correction is needed, because only $q_\tau$ decides whether $\tau$ is excluded. Under full exchangeability in the no-change case, rejecting when $n\notin\mathcal C_{1-\alpha}$ gives a level-$\alpha$ global test.

For multiple changes, let $\boldsymbol{\eta} =\{\eta_1,\ldots,\eta_K\}$ and $\hat{\boldsymbol{\eta}}$ be the true and estimated sets of change-points, respectively, and write $\hat{K} = |\hat{\boldsymbol{\eta}}|$. For two nonempty sets, a common measure of location error is the Hausdorff distance
\begin{align}
	&d_H(\hat{\boldsymbol\eta},\boldsymbol\eta) \\ 
	&=\max\{\max_{s\in\hat{\boldsymbol\eta}}\min_{t\in\boldsymbol\eta}\lvert s-t\rvert,\ \max_{t\in\boldsymbol\eta}\min_{s\in\hat{\boldsymbol\eta}}\lvert s-t\rvert\}.  \notag
	\label{eq:hausdorff}
\end{align}
The two terms measure the worst distance from an estimate to its nearest true change and from a true change to its nearest estimate, respectively. For example, if the true change is at 50 and the estimated changes are at 49 and 51, the Hausdorff distance is only one observation. The estimated number of changes is nevertheless incorrect.

Change-point count consistency, $\Pp(\widehat K=K)\to1$, set localization, and false-change control address different errors. Penalized criteria can select the number of changes \citep{yao1988}, and explicit error-rate methods can control overestimation \citep{chen2023multiple}. Interval choice matters because nearby changes in opposite directions, such as a mean increase followed by a decrease, can cancel in a contrast computed over a long interval. To isolate individual changes, wild binary segmentation samples intervals at random \citep{fryzlewicz2014}, and seeded binary segmentation uses a deterministic multiscale collection \citep{kovacs2023}. Narrowest-over-threshold and multiscale methods also exploit local information \citep{frick2014,baranowski2019}. Binary segmentation and its interval-sampling variants work top-down, whereas bottom-up methods start from short segments and assess local evidence before merging adjacent segments \citep{maeng2026high, park2026multiple}. Guarantees for global penalized methods depend on the segment costs and the penalty \citep{killick2012,maidstone2017,arlot2019}.

\citet{pilliat2023} connect local tests to segmentation through two requirements that must hold simultaneously: all tests on homogeneous windows fail to reject, and for each sufficiently strong change some suitably located window rejects. On this event, their aggregation localizes those changes at the corresponding scales and satisfies their criterion for avoiding spurious estimates. Their minimax result for sparse Gaussian mean changes uses model-specific local tests and signal-to-spacing conditions.

\begin{table*}[t]
\caption{Statistical guarantees in change-point analysis. Each row gives the formal target, the main requirements, and related claims that the guarantee does not imply. $\Pp_0$ and $\Pp_1$ denote probabilities under the null and the alternative. $\Theta$ indexes the data-generating distributions in the model.}
\label{tab:guarantees}
\centering
\scriptsize
\begin{tabularx}{\textwidth}{@{}p{0.16\textwidth}p{0.24\textwidth}p{0.27\textwidth}Y@{}}
\toprule
Claim & Formal target & Main requirements & Does not imply \\
\midrule
Population identification & $\Delta_{\mathcal R}(P,Q)=0$ only if $P=Q$ & Injective embedding of probability distributions, strong negative type, or identifiable classification target & Useful finite-sample power or consistency when dimension grows \\
\specialrule{0.3pt}{1.1pt}{1.1pt}
Fixed-split size & $\Pp_0\{S_n(t)>c_t\}\leq\alpha$ for fixed $t$ & Null distribution of $S_n(t)$, or valid resampling, at that split & Size control after maximizing over $t$ \\
\specialrule{0.3pt}{1.1pt}{1.1pt}
Scan-level size & $\Pp_0\{\max_tS_n(t)>c_n\}\leq\alpha$ & Functional or maximal approximation, or full-scan resampling & Validity after repeated segmentation or online reuse \\
\specialrule{0.3pt}{1.1pt}{1.1pt}
Detection consistency & $\Pp_1\{\max_t S_n(t) > c_n\} \to 1$ & Signal exceeds the calibrated threshold and dominates the maximal stochastic error & Consistent change count or a localization rate \\
\specialrule{0.3pt}{1.1pt}{1.1pt}
AMOC localization & $|\widehat\tau-\tau|=O_{\Pp}(r_n)$ & Uniform concentration and a population margin & Exact recovery or false-change control \\
\specialrule{0.3pt}{1.1pt}{1.1pt}
Location confidence set & $\inf_{\theta\in\Theta}\Pp_\theta\{\tau\in\mathcal{C}_{1-\alpha}\}\ge 1-\alpha$   & Candidate-wise valid tests and justified inversion or simultaneous construction & A short set, detection power, or post-segmentation simultaneity \\
\specialrule{0.3pt}{1.1pt}{1.1pt}
Change-count consistency & $\Pp(\widehat K=K)\to1$ & Detection of every change and a penalty or threshold that rejects spurious splits & A set-localization rate or finite-sample error control \\
\specialrule{0.3pt}{1.1pt}{1.1pt}
Set localization & $d_H(\hat{\boldsymbol\eta},\boldsymbol\eta)=O_{\mathbb P}(r_n)$  & Isolation of every change and control of missed and extra estimates & Correct change count or exact recovery \\
\specialrule{0.3pt}{1.1pt}{1.1pt}
False-change control & Finite-sample family-wise or false-discovery target & Simultaneous or selective calibration of the full segmentation & Change-count consistency or a localization rate \\
\specialrule{0.3pt}{1.1pt}{1.1pt}
Online false-alarm guarantee & ARL, error over a finite horizon, or anytime error & Stopping-time-specific calibration & Anytime error control (from an ARL bound alone) \\
\bottomrule
\end{tabularx}
\end{table*}

%%%%%%%%%%%%%%%%%%%%%%%%%%%%%%%%%%%%%%%%%%%%%

\subsection{Computation and statistical guarantees}

Storing a full Gram or distance matrix requires $O(n^2)$ memory. Exact pruning can reduce the number of candidates evaluated while preserving an optimizer of the specified objective \citep{killick2012, maidstone2017}. Random features \citep{rahimi2007random} and incomplete $U$-statistics \citep{li2015mstat} approximate local scores, and interval subsampling \citep{fryzlewicz2014, baranowski2019, kovacs2023} changes which intervals are searched. On a fixed set of candidate splits, a uniform bound on the score approximation error can be added to the error bound $a_n$ used in the detection and localization arguments around~\eqref{eq:argmaxmargin}. For methods that rely on isolating individual changes, interval subsampling also requires control of the probability that suitable intervals are retained. The critical value must then be computed for the procedure as implemented, including its scores, search, and randomization.

The Reliever method of \citet{qian2025reliever} fits models on a predetermined collection of proxy intervals. It evaluates the loss on each candidate segment using that segment's observations and a model fitted on a proxy interval contained in it. Several candidate segments can thus reuse the same fit. Combined with optimal partitioning in the high-dimensional linear regression model they study, Reliever achieves rates for change-point localization and coefficient estimation that are optimal up to logarithmic factors. Extending this idea to learned distributional contrasts would require controlling how much the reused fits change the contrasts over the intervals searched.

%%%%%%%%%%%%%%%%%%%%%%%%%%%%%%%%%%%%%%%%%%%%%%%%%%%%%%%%%%%%%%%%%%%%%%%%%%%%%%%%%%%%%%%%%%%%%%%%%%%%%%%%%%%%%%%%%%%%%%%%%%%%%%%%%%%%%%%%%%%%%%%%

\section{Choosing a change-point method} \label{sec:diagnostics}

This section turns the framework of Sections \ref{sec:framework} and \ref{sec:guarantees} into two questions to settle before choosing a method: what signal the representation retains (Section \ref{subsec:retain}) and what the analysis should establish (Section \ref{subsec:establish}).

%%%%%%%%%%%%%%%%%%%%%%%%%%%%%%%%%%%%%%%%%%%%%

\subsection{What signal does the representation retain?} \label{subsec:retain}

Two mean shifts of the same size can favor different scans. Suppose $P=N_p(0,I_p)$ and $Q=N_p(\delta,I_p)$. If only the first coordinate increases by $a>0$, the largest absolute coordinate shift is $a$, while the mean shift along the unit vector $u=(1,\ldots,1)^\top/\sqrt p$ is $a/\sqrt p$. If every coordinate increases by $a/\sqrt p$, the shift has the same Euclidean length, but the largest coordinate shift is now $a/\sqrt p$ and the shift along $u$ is $a$. A shift orthogonal to $u$ leaves the entire projected distribution unchanged. When the direction is estimated from the sequence, its alignment with the mean shift also affects localization accuracy \citep{wang2018inspect}.

A mean-based scan can also respond to a change in scale. Even when both populations have mean zero, a variance change can alter the sampling variability of the difference between sample means and hence the probability of rejection; in Figure \ref{fig:representation-power}, Max-CUSUM had power 0.378 against the isotropic scale change at $p = 200$. This sensitivity does not by itself make the scan consistent against scale changes.

\begin{example}[A change in correlation with fixed marginals] \label{ex4.1}
Suppose each observation $(U_i,V_i)$ is jointly Gaussian with standard normal marginals. Its correlation changes from zero to $\kappa\ne0$. Either coordinate alone has the same distribution throughout, so Proposition~\ref{prop:blindness} applies with $R(u,v) = u$ or $R(u,v) = v$, and each marginal discrepancy in~\eqref{eq:coordinate-signal} is zero. The product $U_{i}V_{i}$ of each observation gives a scalar feature with mean zero before the change and $\kappa$ afterwards. As noted in Section \ref{subsec:coordinate}, the dependence is also visible in the joint distribution of the two sample means. In a block of $m$ independent observations from either regime, the two coordinate means are each distributed as $N(0,1/m)$, but their covariance is zero before the change and $\kappa/m$ afterwards.
\end{example}

%%%%%%%%%%%%%%%%%%%%%%%%%%%%%%%%%%%%%%%%%%%%%

\subsection{What should the analysis establish?} \label{subsec:establish}

In the MIT contact study, each day contributes one network. If the question concerns how many distinct pairs meet, the daily count measures the quantity of interest directly. If it matters who meets whom, the Frobenius distance between graph Laplacians, with participants in the same order each day, distinguishes networks with equal counts but different pairs. That distance also responds to changes in the number of pairs, so a peak in the network scan alone leaves open which feature changed.

Suppose the aim is to test for a change-point during the academic term. Permuting the days freely would require them to be exchangeable under the null. A weekly rhythm or persistence from one day to the next can violate that assumption even in the absence of a change-point. Calibration must account for the search over candidate dates and for any choice between the count and network scans made after inspecting their peaks. If the aim is to give a confidence set for the date, its coverage needs a separate justification. When several changes are expected, missed or spurious change-points may matter most (Section \ref{subsec:location}); in online monitoring, detection delay must be assessed under a specified false-alarm criterion (Section \ref{subsec:monitoring}).

Beyond the aim, the available data and the computing budget also narrow the choice. In a high-dimensional sequence, a projection may retain a sparse shift, yet the shortest segment may provide too few observations to estimate its direction reliably. A guarantee derived for independent, light-tailed observations may not apply when extreme values cluster in time. If computing all pairwise distances or repeatedly fitting a classifier is too costly, both the calibration and any claim about location accuracy should be based on the approximation actually used.

%%%%%%%%%%%%%%%%%%%%%%%%%%%%%%%%%%%%%%%%%%%%%%%%%%%%%%%%%%%%%%%%%%%%%%%%%%%%%%%%%%%%%%%%%%%%%%%%%%%%%%%%%%%%%%%%%%%%%%%%%%%%%%%%%%%%%%%%%%%%%%%%

\section{Calibration, power, and localization in simulations}
\label{sec:protocol}

A scan can have high power and still place the change far from its true location. To see when these outcomes diverge, we compare five scans on the same sequences of vectors, with common candidate splits and method-specific critical values. We then examine what changes when observations are serially dependent or when each observation is itself an estimated covariance matrix.

%%%%%%%%%%%%%%%%%%%%%%%%%%%%%%%%%%%%%%%%%%%%%

\subsection{Scan statistics and simulation design} \label{subsec:design}

Each Euclidean sequence has $n=120$ independent observations of dimension $p\in\{5,50,200\}$. Under an alternative, the change occurs at $\tau=60$. All five scans maximize a score over the candidate splits $\mathcal T_n=\{24,\ldots,96\}$, which corresponds to trimming at $\epsilon = 0.2$ in Section \ref{subsec:fixed}. The first two use the coordinate CUSUM vector
\begin{equation}
C_t=\sqrt{\frac{t(n-t)}{n}}\left(\overline X_{1:t}-\overline X_{(t+1):n}\right).
\label{eq:empirical-cusum}
\end{equation}
Max-CUSUM takes the largest absolute coordinate, $\max_j|C_{t,j}|$. The dense projection takes $|\boldsymbol 1^\top C_t|/\sqrt p$, the absolute contrast along the prespecified all-ones direction.

The energy and Gaussian MMD scans use Euclidean distance and the kernel $\exp\{-\|x-y\|^2/(2h^2)\}$, respectively. In each sequence, $h^2$ is the median of the positive squared distances between observations in the pooled sample. Both scans omit diagonal terms from the within-block averages and weight the two-sample contrast by $t(n-t)/n$. The MMD score thus uses the $U$-statistic form of squared MMD with a bandwidth chosen from the data. The graph scan uses a graph that connects two observations if either is among the other's five nearest neighbors. We call this the cross-edge scan. At each candidate split it subtracts the observed number of edges crossing the split from the graph-conditional permutation mean and divides by the permutation standard deviation. A large positive score indicates a deficit of crossing edges.

Under the null, all observations are drawn from $N(0,I_p)$. Table~\ref{tab:simulation-design} gives the post-change distributions under the alternatives. For the radial mixture, each post-change observation is generated as $SZ$, where $Z\sim N(0,I_p)$ and $S$ is independent of $Z$. The scale $S$ equals 0.65 with probability 0.8 and $b$ otherwise, where $b^2=(1-0.8\times 0.65^2)/0.2$. A single draw of $S$ scales all coordinates of an observation. Since $\E S^2=1$, $Z$ and $SZ$ have the same mean and covariance but different distributions of radii.
\begin{table*}[h!]
\caption{Euclidean alternatives. Each sequence changes after observation 60 from $N(0,I_p)$ to the indicated distribution.}
\label{tab:simulation-design}
\centering
\scriptsize
\begin{tabularx}{\textwidth}{@{}p{0.18\textwidth}p{0.37\textwidth}p{0.20\textwidth}Y@{}}
\toprule
Change & Post-change distribution & Unchanged from baseline & Feature carrying the change \\
\midrule
Sparse mean & $N(0.80e_1,I_p)$ where $e_1=(1,0,\ldots,0)^\top\in\mathbb R^p$ & Covariance & Mean of the first coordinate \\ \specialrule{0.3pt}{1.1pt}{1.1pt}
Dense mean & $N(0.65\boldsymbol 1/\sqrt p,I_p)$ & Covariance & Mean along the all-ones direction \\ \specialrule{0.3pt}{1.1pt}{1.1pt}
Isotropic scale & $N(0,1.25^2 I_p)$ & Mean and spherical symmetry & Scale in every coordinate \\ \specialrule{0.3pt}{1.1pt}{1.1pt}
Correlation only & Centered Gaussian with unit variances, pairwise correlation 0.60 within the first five coordinates and zero elsewhere & All univariate marginals & Dependence within a fixed set of coordinates \\ \specialrule{0.3pt}{1.1pt}{1.1pt}
Radial mixture & $SZ$ with $\E S^2=1$ as defined in Section \ref{subsec:design} & Mean and covariance & Distribution of the radius \\
\bottomrule
\end{tabularx}
\end{table*}

For each dimension and method, we estimate the critical value as the empirical 0.95 quantile of the scan maximum from 10,000 independent sequences generated under the specified Gaussian null. We use these model-based critical values so that all scans are compared at the same nominal level. We estimate the type I error rate from 2,000 independent null sequences and the power against each alternative from 500 sequences shared across methods, so comparisons between methods are paired. Pointwise Wilson intervals describe Monte Carlo uncertainty in each estimated rate, conditional on the estimated critical value. Section \ref{subsec:scan} examines detection and localization for the radial mixture and the sparse mean shift at $p = 200$, reporting the miss rate (one minus the power), the median absolute location error among detections, and the mean of an illustrative loss that assigns cost $n$ to a missed change and cost $|\widehat\tau-\tau|$ to a detected change.

%%%%%%%%%%%%%%%%%%%%%%%%%%%%%%%%%%%%%%%%%%%%%

\subsection{Calibrating the search and choosing the permutation unit}
\label{sec:simulation-calibration}

\begin{figure*}[h!]
	\centering
	\includegraphics[width=\textwidth]{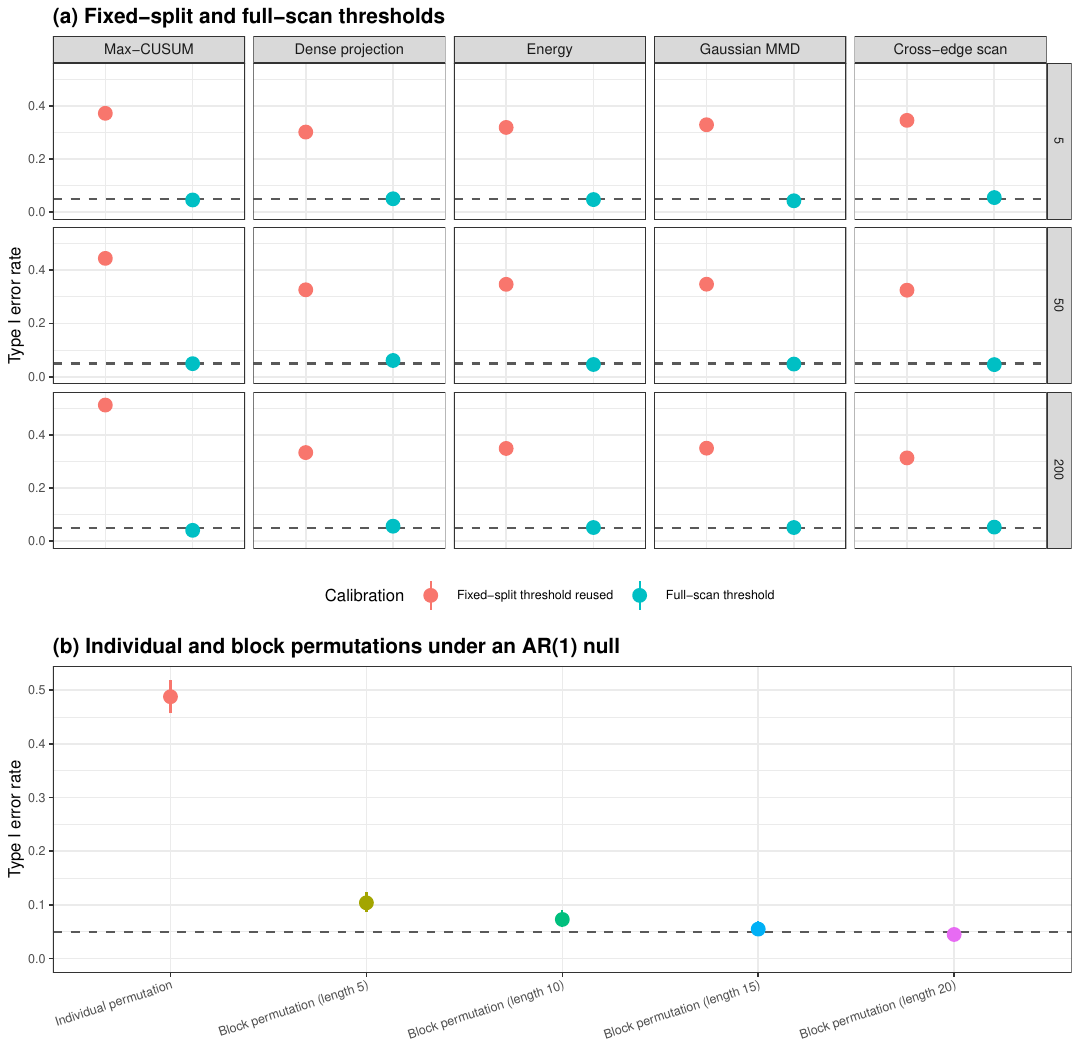}
	\caption{Type I error rates. Panel (a) compares fixed-split and full-scan thresholds applied to the scan maximum on 2,000 independent Gaussian null sequences. Rows give $p$ and columns give the scan. Panel (b) compares individual and block permutations on 1,000 Gaussian AR(1) null sequences. Vertical bars give 95\% Wilson intervals and dashed lines mark the nominal level 0.05.}
	\label{fig:calibration}
\end{figure*}

A threshold calibrated for the prespecified split $t=60$ gave type I error rates of 0.042--0.057 at that split across the 15 method--dimension combinations. Applying the same thresholds to the maximum over all candidate splits raised the type I error rate to 0.301--0.512 on the same sequences. Calibrating each maximum against null maxima from the same search brought the type I error rate back to 0.041--0.061 (Figure~\ref{fig:calibration}).

Dependence restricts which rearrangements of the observations are valid. We generated univariate Gaussian AR(1) sequences of length 120 with coefficient 0.5 and stationary variance one, discarding the first 200 observations as burn-in. For each sequence and permutation scheme, we recomputed the absolute CUSUM scan (\ref{eq:empirical-cusum}) with $p = 1$ over the same candidate splits for 199 random permutations and computed the permutation $p$-value. The test that permutes individual observations rejected 488 of 1,000 null sequences (95\% Wilson interval 0.457--0.519). Permuting intact, nonoverlapping blocks of lengths 5, 10, 15, and 20 gave type I error rates of 0.104, 0.073, 0.055, and 0.045, respectively. When we added 0.85 to observations 61--120, the power estimated from 500 sequences was 0.704, 0.562, 0.406, and 0.220 for the same block lengths. Individual permutation had power 0.948 against this alternative, but its type I error rate was already 0.488. In this design, the estimated type I error rate moved closer to the nominal level as the block length increased, while the power decreased. Longer blocks preserve more dependence within each block but leave fewer blocks to rearrange--only six when the block length is 20. These results illustrate sensitivity to block length; they do not establish finite-sample validity of block permutation.

%%%%%%%%%%%%%%%%%%%%%%%%%%%%%%%%%%%%%%%%%%%%%

\subsection{Which changes do the scans detect and localize?} \label{subsec:scan}

\begin{figure*}[h!]
	\centering
	\includegraphics[width=\textwidth]{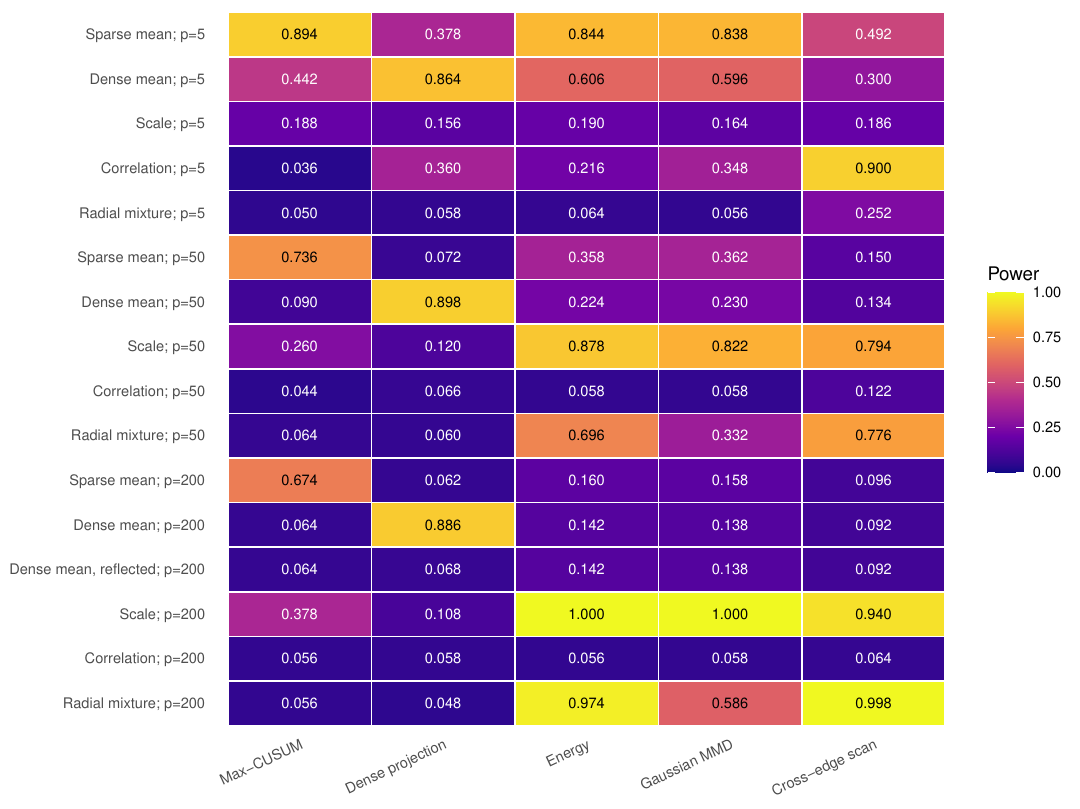}
	\caption{Power under the alternatives of Table \ref{tab:simulation-design}, from 500 simulated sequences per cell. Conditional on the critical values estimated under the Gaussian null, each estimate has a pointwise 95\% Wilson interval with half-width at most 0.044. At $p=200$, the reflected dense alternative is the dense alternative with the signs of half of the coordinates reversed.}
	\label{fig:representation-power}
\end{figure*}

For the sparse mean shift of norm 0.80, the power of Max-CUSUM fell from 0.894 to 0.736 to 0.674 as $p$ increased from 5 to 50 to 200. The corresponding power for the energy scan was 0.844, 0.358, and 0.160 (Figure~\ref{fig:representation-power}). The dense shift has norm 0.65 at every dimension and points along the all-ones direction. At $p=200$, the projection onto that direction detected 443 of 500 changes. Reversing the signs of the last 100 coordinates throughout each of these 500 sequences makes the mean shift orthogonal to the projection while preserving its norm and the noise distribution. The projection scan then detected only 34 changes, a paired difference in power of 0.818 (Monte Carlo standard error 0.017). The reflection preserves all pairwise distances and absolute coordinate contrasts, so the other four scans gave identical decisions within every pair. Here all 200 coordinates change by the same absolute amount in both alternatives. Their signs determine whether the fixed projection retains the shift.

The effect of dimension depended on which coordinates carried the signal. When correlations changed only within the first five coordinates, the power of the cross-edge scan fell from 0.900 to 0.122 to 0.064 as $p$ grew. With an isotropic scale change, every additional coordinate carried signal, and the power of the energy scan rose from 0.190 to 0.878 to 1.000. For the radial mixture it rose from 0.064 to 0.696 to 0.974 even though the first two moments were unchanged.

For the radial mixture at $p=200$, the cross-edge scan detected 499 of 500 changes. Among detections its median change-point estimate was 34 (interquartile range 33--35), well before the true change-point at 60. At $t=\tau=60$, every simulated sequence had more crossing edges than its permutation expectation, giving a median standardized deficit of $-2.40$. Thus the score at the true change-point had the opposite sign to the one the scan rewards. \cite{chenfriedman2017} explain related failures of the cross-edge deficit under high-dimensional scale alternatives. Our example adds that maximizing this score over time can yield a location estimate far from the true change-point. On the same sequences, the energy scan detected 487 changes with a median absolute error of 2 observations among detections. 

To see whether the choice of score explained this early peak, we kept the graphs and candidate splits fixed and changed only the edge-count score. Let $R_1(t)$ and $R_2(t)$ count edges with both endpoints in the first and second blocks, respectively. We considered the weighted count of \citet{chen2018weighted},
\[
 R_w(t)=\frac{(n-t-1)R_1(t)+(t-1)R_2(t)}{n-2},
\]
and the generalized score $Z_w(t)^2+Z_d(t)^2$ studied by \citet{chu2019}. Here $Z_w$ and $Z_d$ are $R_w$ and $R_1-R_2$, respectively, centered and scaled by their graph-conditional permutation means and standard deviations. We scanned $Z_w$ for a one-sided increase. The generalized score also responds to imbalance between within-block edge counts in either direction \citep{chenfriedman2017}. Each score had its own estimated critical value from full scans on the same 10,000 null sequences.

At $t=\tau=60$, where the two blocks have equal size, $R_w$ equals half the total number of edges minus half the number of crossing edges. Its standardized value therefore equals the original standardized deficit there. The one-sided weighted scan detected only 2 of the 500 changes. The generalized scan detected all 500, with a median change-point estimate of 61 (interquartile range 60--62) and a median absolute error of 1. At $t=\tau$, the median $Z_d$ was $-5.93$, an imbalance that contributes through its square. On the 2,000 independent null sequences, the type I error rates of the cross-edge, weighted, and generalized scans were 0.053, 0.046, and 0.040, respectively.

\begin{figure*}[h!]
	\centering
	\includegraphics[width=.88\textwidth]{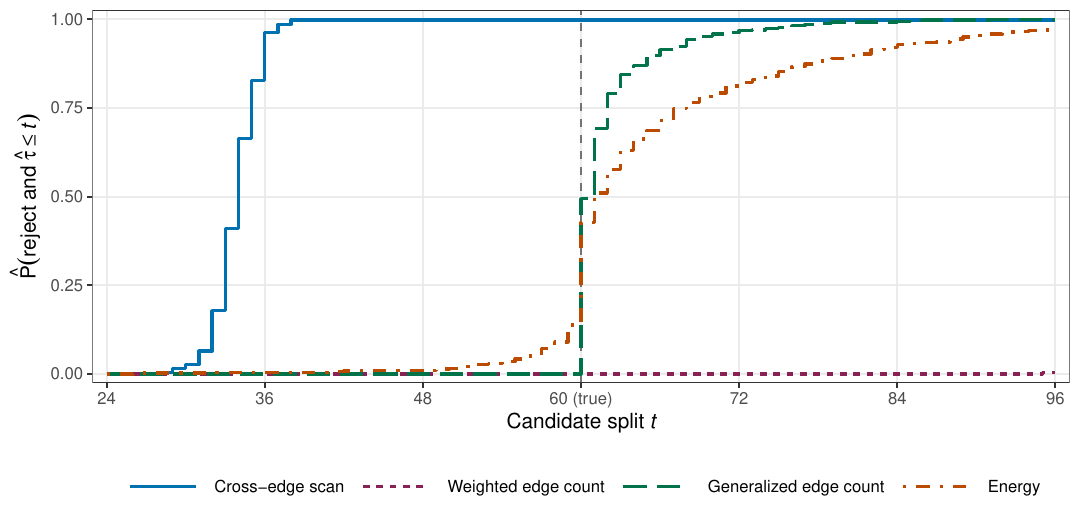}
	\caption{Detection and localization for the same 500 radial-mixture sequences at $p=200$ used in Figure \ref{fig:representation-power}. At each candidate split $t$, the curve gives the fraction of all 500 sequences in which the scan both detects a change and locates it at or before $t$. The right endpoint is the power. The vertical line marks the true change-point at 60.}
	\label{fig:radial-location}
\end{figure*}

Figure~\ref{fig:radial-location} plots the empirical probability of both rejecting and estimating the change at or before $t$. The cross-edge and generalized edge-count curves end at nearly the same height, reflecting similar power, but the cross-edge curve rises well before 60 and the generalized curve near 60. For this mixture and dimension, the same edges can locate the change accurately when the score uses the within-block imbalance. Location error among detections can also hide misses (Section \ref{subsec:location}). Under the sparse shift at $p = 200$, Max-CUSUM had a median absolute error of 2 observations among detections but missed 0.326 of the changes; charging $n = 120$ for each miss gave an average loss of 42.5 observations, compared with 120 for a scan that never rejects.

%%%%%%%%%%%%%%%%%%%%%%%%%%%%%%%%%%%%%%%%%%%%%

\subsection{Spectral and orientation changes in covariance matrices} \label{subsec:spectral}

Suppose each observation is a $3\times3$ sample covariance matrix estimated from $m$ independent Gaussian vectors. The sequence contains $n=80$ such matrices and changes after $\tau=40$. We vary the number of measurements per matrix, $m\in\{12,40\}$. The initial population covariance is $\Sigma_0=\operatorname{diag}(2,1,0.5)$. A spectral change replaces it with $\operatorname{diag}(2.5,0.85,0.40)$. An orientation change replaces it with $W\Sigma_0W^\top$, where $W$ rotates the first two coordinates by $\pi/8$ and leaves the eigenvalues unchanged.

We apply two energy scans to the same sample covariance matrices, adding a ridge of $10^{-8}I$ before taking logarithms. One scan vectorizes the full matrix logarithm, storing the diagonal and one copy of each off-diagonal entry; this gives the log-Euclidean metric of Section \ref{subsec:object}. Scaling each stored off-diagonal entry by $\sqrt2$ makes Euclidean distance between these vectors equal the Frobenius distance between matrix logarithms. The other scan uses Euclidean distances between vectors of sorted log eigenvalues. Both scans search over the candidate splits $\{16, \ldots, 64\}$, the same 20\% trimming as in Section \ref{subsec:design}. For each $m$, we use 2,000 null sequences to calibrate the two scans and 2,000 independent null sequences to evaluate them. Across both scans and both values of $m$, type I error rates ranged from 0.044 to 0.054.

Under the spectral change with $m=12$, the full-matrix scan detected 311 of 500 changes and the eigenvalue scan detected 250. At $m=40$ both detected all 500. Under a rotation, however, the post-change sample covariance is distributed as $W\widehat\Sigma W^\top$, where $\widehat\Sigma$ is distributed as a pre-change sample covariance. Its sorted eigenvalues therefore have exactly the same distribution on both sides of the change for every $m$, so Proposition \ref{prop:blindness} applies with $R$ the map to sorted log eigenvalues. The eigenvalue scan had power 0.046 and 0.032 for $m=12$ and 40, respectively. For the full-matrix scan the corresponding power was 0.724 and 1.000. Increasing the precision of each matrix estimate helped detect the rotation only when the representation retained orientation.

%%%%%%%%%%%%%%%%%%%%%%%%%%%%%%%%%%%%%%%%%%%%%%%%%%%%%%%%%%%%%%%%%%%%%%%%%%%%%%%%%%%%%%%%%%%%%%%%%%%%%%%%%%%%%%%%%%%%%%%%%%%%%%%%%%%%%%%%%%%%%%%%

\section{Dependence, structured observations, and online monitoring}
\label{sec:extensions}

In the autoregressive experiment of Section~\ref{sec:simulation-calibration}, shuffling individual observations destroys the serial correlation allowed under the null. In the covariance experiment of Section \ref{subsec:spectral}, whole matrices are exchangeable under the null when each is computed from the same number $m$ of independent draws from a common distribution. The appropriate unit therefore depends on how the observations were generated and what variation the null allows. In a continuing stream the analyst must also choose how to measure false alarms over time. Table~\ref{tab:inferential-unit} gives examples of these choices.

\begin{table*}[t]
\caption{Inferential targets, observation units, calibration, and common pitfalls across data settings.}
\label{tab:inferential-unit}
\centering
\scriptsize
\begin{tabularx}{\textwidth}{@{}p{0.17\textwidth}p{0.22\textwidth}p{0.18\textwidth}p{0.22\textwidth}Y@{}}
\toprule
Setting & Target & Unit or time index & Basis for calibration & Possible pitfall \\
\midrule
Weakly dependent observations & Marginal distribution of each observation & Individual time point & Long-run variance, dependent multipliers, or block resampling under suitable assumptions & Shuffling individual observations \\ \specialrule{0.3pt}{1.1pt}{1.1pt}
Changes in dependence & Joint distribution at selected lags & Overlapping lag vectors & Joint null distribution of the statistics over the lags and splits searched & Missing changes at omitted lags \\ \specialrule{0.3pt}{1.1pt}{1.1pt}
Tied observations & Distribution including ties & Whole observation & Permutation under exchangeability with tie-aware graph or rank scores & Breaking ties by temporal order \\ \specialrule{0.3pt}{1.1pt}{1.1pt}
Repeated measurements & Change across successive subjects & Subject record & Whole-subject permutation under exchangeability & Shuffling measurements across subjects \\ \specialrule{0.3pt}{1.1pt}{1.1pt}
Object-valued data & Distribution of the measured objects & Whole object & Object-level permutation under exchangeability & Confounding precision with a change in the underlying objects \\ \specialrule{0.3pt}{1.1pt}{1.1pt}
Network snapshots & Edge-probability matrix & Network on a fixed set of vertices & Null model for aligned network snapshots & Ignoring persistent edges \\ \specialrule{0.3pt}{1.1pt}{1.1pt}
Online ARL control & Expected time to a false alarm & Stopping time & Threshold chosen for the target ARL & Reusing an offline test at each time \\ \specialrule{0.3pt}{1.1pt}{1.1pt}
Finite-sample anytime control & Probability of ever raising a false alarm & Entire monitoring period & Valid e-process or summable error budgets & Inferring anytime control from ARL \\
\bottomrule
\end{tabularx}
\end{table*}

%%%%%%%%%%%%%%%%%%%%%%%%%%%%%%%%%%%%%%%%%%%%%

\subsection{Marginal changes and changes in serial dependence}

A change in an autoregressive coefficient can leave the marginal distribution of every observation intact. Consider a Gaussian AR(1) process, initialized with $X_1\sim N(0,1)$, whose coefficient changes from $\beta_0$ to $\beta_1$, with $|\beta_0|,|\beta_1|<1$. Let the Gaussian innovations be independent of the past, with mean zero and variance $1-\beta_0^2$ or $1-\beta_1^2$ according to the current coefficient. Then $X_i\sim N(0,1)$ throughout, while the lag-one covariance changes from $\beta_0$ to $\beta_1$. Thus the joint distribution of $(X_{i-1},X_i)$ records the change. Consecutive lag vectors share an observation and are therefore dependent, which calibration must respect.

The NP-MOJO procedure of \cite{mcgonigle2025} compares empirical joint characteristic functions of observations and their lags in adjacent windows. Its consistency results apply to changes visible at the selected lags under its dimension and dependence assumptions. A change in the distribution of the full process may leave those lag distributions unchanged. For regenerating Markov chains, \citet{banerjee2026} instead compare the stationary marginal distributions through their empirical distribution functions. Regeneration controls the fluctuations of these estimates. A change in the transition kernel that preserves the stationary marginal leaves this population discrepancy at zero, although it may alter the statistic's sampling distribution. For changes in a multivariate density, \citet{madridpadilla2023mixing} establish detection and localization results under mixing conditions.

For graph scans under local dependence, \citet{chen2019local} studies circular block permutation with a randomized starting point. Block length also mattered in the nonoverlapping block permutations of Section~\ref{sec:simulation-calibration}. If the procedure also searches several lags or window lengths, its calibration must cover that search.

%%%%%%%%%%%%%%%%%%%%%%%%%%%%%%%%%%%%%%%%%%%%%

\subsection{Ties and repeated measurements}

Rounding can give distinct observations the same value. When tied distances admit several optimal similarity graphs, \citet{song2022repeated} construct scans by averaging edge counts over the admissible graphs or by computing the counts on their union. Repeated measurements on a subject, such as a patient, raise a different question about the sampling unit. Suppose patients arrive in sequence, each with a longitudinal record. To test for a change across successive patients, whole records can be permuted if the patients are exchangeable under the null. This preserves the order of measurements within each record. For a change within each patient's record, however, permuting whole records does not randomize the within-record time ordering. A valid reference distribution must instead respect the temporal structure within records and any dependence across subjects. Within-record permutations are valid only when the joint null distribution is invariant to the chosen transformations.

%%%%%%%%%%%%%%%%%%%%%%%%%%%%%%%%%%%%%%%%%%%%%

\subsection{Object geometry and measurement precision}

A representation encodes which transformations of an object are treated as irrelevant. The order in which points are listed in a point cloud is arbitrary, so a representation of point clouds should ignore it. In a network on named participants, by contrast, exchanging vertex labels in one snapshot changes which people are recorded as meeting, so the representation must keep the labels. In the covariance experiment, sorted eigenvalues retain the spectrum of a matrix while discarding orientation relative to the measurement axes.

Fr\'echet means and variances measure location and dispersion in a metric space \citep{dubey2020}. For a fuller comparison, \citet{dubeyzheng2026profiles} use distance profiles, the distributions of $\rho(\omega,X)$ at reference objects $\omega$, to scan for a single change and estimate its location. For weakly dependent objects, \citet{zhang2026object} use pairwise distances in a self-normalized statistic and establish consistency of a wild binary segmentation procedure under their assumptions (see Section \ref{subsec:object}).

When object $i$ in a sequence of $n$ objects is estimated from $m_i$ measurements, the observed sequence reflects both differences among the underlying objects and variation due to measurement. For covariance matrices, for example, a change in $m_i$ alters the distribution of the sample covariance matrices even when the population covariance is constant. Attributing a detection to a change in the underlying covariance therefore requires accounting for the numbers of measurements $m_i$. Resampling measurements within an object can assess its estimation uncertainty under an appropriate sampling model. A permutation across time moves whole estimated objects and requires their exchangeability under the null. This exchangeability generally fails when the measurement counts $m_i$ vary systematically over time and alter the distributions of the estimates, even if the underlying objects do not change.

For sequences of networks, \citet{wang2021network} study independent adjacency matrices on common vertices with independent Bernoulli edges above the diagonal and piecewise-constant edge probabilities. The common vertex labels allow the same edge to be averaged across snapshots. Under their signal and rank conditions, a local refinement step attains a minimax localization rate. Edge persistence across time would require a dependence model to justify a comparable rate.

%%%%%%%%%%%%%%%%%%%%%%%%%%%%%%%%%%%%%%%%%%%%%

\subsection{False alarms in online monitoring} \label{subsec:monitoring}

In online monitoring, the analyst must decide whether to raise an alarm using only the observations available so far. The time of the first alarm is a stopping time $T$. Classical sequential theory relates false alarms to detection delay \citep{page1954,lorden1971,tartakovsky2014}. Methods for distributional changes include nearest-neighbor and graph detectors \citep{chen2019sequential,chu2022sequential} and NP-FOCuS, which tracks the proportions of observations below selected thresholds and uses functional pruning \citep{romano2024}. \citet{wang2024sequential} review the computational costs of sequential procedures.

Let $\Pp_\infty$ and $\E_{\infty}$ denote probability and expectation when no change occurs. Three common false-alarm criteria are
\begin{equation}
\begin{aligned}
\E_\infty T&\geq\gamma,\\
\Pp_\infty(T\leq H)&\leq\alpha_H,\\
\Pp_\infty(T<\infty)&\leq\alpha.
\end{aligned}
\label{eq:onlinecriteria}
\end{equation}
The first sets a minimum average run length (ARL). The second bounds the probability of a false alarm by a specified horizon $H$, and the third bounds that probability over the entire stream. For example, independent tests that each signal with probability $1/\gamma$ give a geometric waiting time with mean $\gamma$ and eventually signal with probability one. Such a procedure meets the first criterion with equality yet violates the third for every $\alpha < 1$. Under the third criterion, $\alpha<1$ gives positive probability to $T=\infty$, so $\E_{\infty}T = \infty$ and $\Pp_{\infty}(T \le H) \le \alpha$ for every $H$; the third criterion thus implies the first for every $\gamma$ and the second with $\alpha_H = \alpha$. \citet{aue2024monitoring} review nonparametric CUSUM procedures that use a stable historical sample. Their monitoring boundaries control the probability of a false alarm over the chosen horizon asymptotically as the historical sample grows.

The way a procedure accumulates evidence affects which bound it satisfies. The e-Shiryaev--Roberts and e-CUSUM detectors aggregate evidence processes started at candidate change times. With threshold $1/\alpha$, both have a nonasymptotic ARL of at least $1/\alpha$ under the pre-change class \citep{shin2024edetectors}. By contrast, a single null e-process with initial value one (a nonnegative process whose expected value at any stopping time is at most one under the null) reaches $1/\alpha$ with probability at most $\alpha$. This anytime guarantee does not automatically extend to an unadjusted maximum or sum of processes restarted at successive candidate change times; additional weighting or error-budget allocation is needed. A different construction compares forward and backward confidence sequences for the parameter or functional of interest and signals when they cease to intersect \citep{shekhar2023}.

The online kernel CUSUM procedure of \citet{wei2026online} compares recent windows with a historical reference sample. They approximate its ARL and expected detection delay, relating the largest window length to the logarithm of the target ARL. With the window design fixed, the cost per update stays bounded as monitoring continues. Once a rolling reference and the current window contain only observations from the same post-change distribution, their population discrepancy is zero. A detector that retains no earlier evidence may then miss the change.

%%%%%%%%%%%%%%%%%%%%%%%%%%%%%%%%%%%%%%%%%%%%%%%%%%%%%%%%%%%%%%%%%%%%%%%%%%%%%%%%%%%%%%%%%%%%%%%%%%%%%%%%%%%%%%%%%%%%%%%%%%%%%%%%%%%%%%%%%%%%%%%%

\section{Open questions about representations and guarantees}
\label{sec:open}

We discuss four open problems: the cost of choosing representations from data, detection boundaries in growing dimensions, uncertainty in estimated change locations, and late changes under anytime false-alarm control.

%%%%%%%%%%%%%%%%%%%%%%%%%%%%%%%%%%%%%%%%%%%%%

\subsection{The cost of choosing a representation} \label{subsec:cost}

In Section \ref{subsec:scan}, reflecting half of the coordinates removed the mean shift from a prespecified projection. Estimating the direction from the observations might recover the signal, but the same data would then determine both the direction and the reported change-point. More generally, suppose a scan searches over kernel bandwidths or projection directions as well as candidate change times. Under exchangeability, repeating both searches on each permuted sequence calibrates the global test. How much stronger must a change be to attain the location accuracy of a scan whose representation was chosen with knowledge of the alternative? The answer should account for how strongly the candidate scans are correlated under the null and how sharply each population score peaks at the true change. Existing results address parts of this question: ART gives finite-sample rank calibration for scores fitted symmetrically to the pooled data \citep{cui2026art}, and CUSUM-Net relates error in a learned critic to localization accuracy under smoothness assumptions \citep{li2026cusumnet}.

%%%%%%%%%%%%%%%%%%%%%%%%%%%%%%%%%%%%%%%%%%%%%

\subsection{Detection boundaries as dimension grows} \label{subsec:boundary}

In the correlation experiment of Section \ref{subsec:scan}, increasing the dimension adds unchanged coordinates around a block of five affected ones. In the scale experiment, every added coordinate carries signal. A common model for comparing methods could allow a block of $q$ out of $p$ coordinates to change and vary the block size and the magnitude of the change separately. With the type I error rate and candidate splits held fixed across scans, how large a change does each method need for reliable detection? Results on kernel power, smooth-density changes, dense alternatives, and dimension reduction address this question under different models \citep{yan2023,padilla2022,drikvandi2025,yu2026kme}. For changes in marginal distributions, a recent preprint combines sums and maxima of rank statistics to adapt to the number of affected coordinates \citep{zhao2026}. These results are not directly comparable: a comparison of detection boundaries would need to express each method's population discrepancy in terms of the same underlying change.

%%%%%%%%%%%%%%%%%%%%%%%%%%%%%%%%%%%%%%%%%%%%%

\subsection{Location uncertainty} \label{subcec:interval}

Suppose the analyst uses the same data to choose intervals, fit a score using time labels, and report a change location in each selected interval. Existing guarantees address different parts of this problem. For independent Gaussian observations with piecewise-constant means, \citet{jewell2022selective} account for the selection of an estimated change-point when testing a local mean contrast. Multiscale rank tests can control the probability that any flagged interval contains no change \citep{cui2026art}. Other procedures address selection of the number of changes \citep{chen2023multiple} or give a conformal location set for a single change with independent, internally exchangeable segments \citep{dandapanthula2026}. Under what signal and spacing conditions can the reported estimates be matched to distinct changes, and how narrow can simultaneous confidence sets for those locations be? The coverage target must specify how to treat intervals containing more than one change and estimates with no corresponding change.

Localization rates also rest on assumptions that structured data can violate. \citet{dubeyzheng2026profiles} derive their localization rate under the assumption that the distance from each fixed reference object to a random observation has a density. For networks formed from contact indicators on a fixed set of participants, however, the distances between daily graph Laplacians take discrete values. Their theorem therefore does not apply directly, although finite-sample permutation calibration remains valid if the daily networks are exchangeable under the null. A location analysis that accommodates this discreteness and any dependence across days could show whether the MIT records distinguish a change-point during finals week from one in a neighboring week.

%%%%%%%%%%%%%%%%%%%%%%%%%%%%%%%%%%%%%%%%%%%%%

\subsection{Detecting a late change in a stream} \label{subsec:late}

Consider the requirement $\Pp_\infty(T<\infty)\leq\alpha$ in~\eqref{eq:onlinecriteria}. One way to meet it is to start a test at each time $s$ with probability at most $\alpha_s$ of ever raising an alarm under the null, where $\sum_s\alpha_s\leq\alpha$. Because the $\alpha_s$ must be summable, they eventually become small, so tests started late require strong evidence to raise an alarm. If the distribution changes just after time $\tau$, one measure of detection delay is the probability of raising an alarm by $\tau+d$, given no false alarm by $\tau$. How does this probability vary with the arrival time and the size of the change when kernels or graphs are fitted from serially dependent observations? Comparing fixed and fitted representations under the same pre- and post-change distributions and false-alarm bound would help quantify the cost of adaptation for late changes.
\

%%%%%%%%%%%%%%%%%%%%%%%%%%%%%%%%%%%%%%%%%%%%%%%%%%%%%%%%%%%%%%%%%%%%%%%%%%%%%%%%%%%%%%%%%%%%%%%%%%%%%%%%%%%%%%%%%%%%%%%%%%%%%%%%%%%%%%%%%%%%%%%%

\section{Discussion}
\label{sec:discussion}

For a distance of negative type, the energy scan and its induced-kernel counterpart have proportional scores at every candidate split when their estimators and split weights match. They therefore have the same maximizing splits, the same change-point estimates under a common tie-breaking rule, and the same rejection decisions when the critical values are scaled accordingly. In the radial-mixture experiment, by contrast, the cross-edge and generalized edge-count scans both detected nearly every change on the same graph (Figure~\ref{fig:radial-location}), yet the cross-edge deficit typically peaked well before the true change-point. The imbalance between the two within-block edge counts helped the generalized score locate the change. Comparing power alone would have missed this difference.

For applied work, the framework suggests four checks. State the change of scientific interest and confirm that the representation retains it, since no calibration can recover a change that the representation hides (Proposition \ref{prop:blindness}). Calibrate the full search, including any data-driven choice of representation or tuning. Choose the permutation or resampling unit from how the data were generated. Finally, report misses alongside location error, and state which guarantee in Table \ref{tab:guarantees} the reported result carries.

Applied to the MIT contact study, these checks suggest a second analysis. The network scan of \citet{dubeyzheng2026profiles} placed the change on December~15, during finals week. Scanning the daily count of distinct pairs over the same candidate dates would show whether that week also stands out when the identities of the pairs are discarded. In either analysis, evidence for the change-point depends on calibrating the search against a model for ordinary variation across days, including any weekly rhythm.

%%%%%%%%%%%%%%%%%%%%%%%%%%%%%%%%%%%%%%%%%%%%%%%%%%%%%%%%%%%%%%%%%%%%%%%%%%%%%%%%%%%%%%%%%%%%%%%%%%%%%%%%%%%%%%%%%%%%%%%%%%%%%%%%%%%%%%%%%%%%%%%%

\section*{Funding}

This work was supported, in part, by the National Research Foundation of Korea (NRF) grants (RS-2026-25471551, RS-2026-25517529) and the National IT Industry Promotion Agency (NIPA) grant (RS-2026-25621650), both funded by the Korea government (MSIT).

%%%%%%%%%%%%%%%%%%%%%%%%%%%%%%%%%%%%%%%%%%%%%%%%%%%%%%%%%%%%%%%%%%%%%%%%%%%%%%%%%%%%%%%%%%%%%%%%%%%%%%%%%%%%%%%%%%%%%%%%%%%%%%%%%%%%%%%%%%%%%%%%

\bibliographystyle{spr-ims-nameyear}
\bibliography{references}

\end{document}